# Dark matter, dark energy and gravitational proprieties of antimatter

Dragan Slavkov Hajdukovic
PH Division CERN
CH-1211 Geneva 23
dragan.hajdukovic@cern.ch

Abstract
We suggest that the eventual gravitational repulsion between matter and antimatter may be a key for understanding of the nature of dark matter and dark energy. If there is gravitational repulsion, virtual particle-antiparticle pairs in the vacuum, may be considered as gravitational dipoles. We use a simple toy model to reveal a first indication that the gravitational polarization of such a vacuum, caused by baryonic matter in a Galaxy, may produce the same effect as supposed existence of dark matter. In addition, we argue that cancellation of gravitational charges in virtual particle-antiparticle pairs, may be a basis for a solution of the cosmological constant problem and identification of dark energy with vacuum energy. Hence, it may be that dark matter and dark energy are not new, unknown forms of matter-energy but an effect of complex interaction between the quantum vacuum and known baryonic matter.

## 1. Introduction

According to the current concordance model of cosmology, the Earth, the stars, the Galaxies and everything we are familiar with (i.e. everything made from quarks and leptons as building blocks), make up less than 5% of the total matter and energy in the Universe. The rest of 95% is dark energy and dark matter; roughly in proportion 3:1. Hence, the present day Universe is dominated by its "dark side" (it reminds me the human society). However no one knows what dark energy and dark matter are. Their nature is the greatest mystery of contemporary cosmology. We are living in the age of "precision cosmology", i.e. in the time of the unprecedented precision of observations; we are able to observe what's happening but without understanding why it's happening.

The simplest evidence for unseen "dark" matter comes from observation of spiral galaxies. A spiral galaxy is a disk of dust and stars (typically $10^{11}$ stars) rotating about a central nucleus. The velocity $v(r)$ of rotation of stars in the galaxy can be approximately determined as a function of distance $r$ from its centre. The surprise coming from observations is that, outside a radius that contains the most of the visible mass of the galaxy, $v(r)$ remains approximately constant as far out as can be measured. But, if we have to trust the well established law of gravitation, one would expect the velocity $v(r)$ at a radius $r$ to be related to the mass $M(r)$ interior to that radius by a relation roughly like

$$\frac{GM(r)}{r^2} = \frac{v^2(r)}{r} \quad i.e. \quad v(r) = \sqrt{\frac{GM(r)}{r}} \tag{1}$$

Thus, there is a big conflict between expectations that $v(r)$ fall off as $r^{-1/2}$ and observation that $V(r)$ is nearly constant. One possible solution is ad hoc assumption that visible matter of the Galaxy is surrounded by a spherical halo of some yet unknown and invisible "dark" matter. If we want the second of Equations (1) to agree with observations, the mass of dark matter $M_{dm}(r)$ interior to a radius $r$ must be proportional to that radius. Hence, the radial mass distribution $M_{dm}(r)$, and radial density profile $\rho_{dm}(r)$ of dark matter are characterized with proportionalities:

$$M_{dm}(r) \sim r; \quad \rho_{dm}(r) \sim r^{-2} \tag{2}$$

Dark energy was invoked after the recent discovery [1], that the expansion of the universe is accelerating rather than slowing down. As well known [2], the easiest way for producing theoretical models with accelerated expansion of the universe is to assume a positive cosmological constant $\Lambda$ in Einstein's field equations. In fact, in addition to the usual source term, proportional to the energy-momentum tensor $T_{\mu\nu}$, Einstein's equations for the gravitational field can accommodate an additional source term proportional to the metric field $g_{\mu\nu}$ itself:

$$R_{\mu\nu} - \frac{1}{2} g_{\mu\nu} R = \frac{8\pi G}{c^4} T_{\mu\nu} + \Lambda g_{\mu\nu} \qquad (3)$$

where $R_{\mu\nu}$ and $R$ denote the Ricci and scalar curvatures defined with the metric $g_{\mu\nu}$.

The key point is that such an ad hoc introduced cosmological constant has the same physical effect as if an appropriate, constant mass (energy) density is attributed to the vacuum. Equivalently everything that contributes to the mass (energy) density of the vacuum, contributes to the cosmological constant through linear relations:

$$\Lambda = \frac{8\pi G}{c^4} \rho_E \quad \text{and} \quad \Lambda = \frac{8\pi G}{c^2} \rho_m \qquad (4)$$

where $\rho_E$ and $\rho_m$, are constant energy density and constant mass density of the vacuum. At first sight, it looks like a welcome "meeting point" between General Relativity and Quantum Field Theories (QFT). In fact, from the point of view of QFT, the vacuum is not just an empty space as in non-quantum theories, but a still poorly understood "kingdom" of virtual particle-antiparticle pairs and fields. Hence, in principle, QFT can provide a mechanism producing the vacuum energy density and estimate the numerical value of such density. But what a catastrophe! The confrontation of equation (1) with observations [3], places an upper bound on $\Lambda$, and consequently on $\rho_E$ and $\rho_m$:

$$|\Lambda| < 10^{-52} m^{-2} \; ; \; |\rho_m| < 10^{-26} kg/m^3 \; ; \; |\rho_E| < 10^{-9} J/m^3 \qquad (5)$$

while QFT predict values [2] which are, minimum a few tens orders of magnitude larger than these observed values. This dramatic discrepancy (called cosmological constant problem) is the key obstacle for attractive identification of dark energy with the vacuum energy.

In the present paper we present a radically new idea that dark matter and dark energy are not new, unknown forms of matter-energy but an effect of complex interaction between the quantum vacuum and known baryonic matter. Our speculations are based on the conjecture, that there is gravitational repulsion between particles and antiparticles.

## 2. Conjecture of gravitational repulsion between particles and antiparticles

The gravitational proprieties of antimatter are still not known. The AEGIS experiment [4], recently approved at CERN would be presumably the first one in the human history to measure acceleration of antimatter in the gravitational field of the Earth. In fact it would measure the gravitational acceleration of atoms of the antihydrogen, using interferometry; a technique well established for atoms of ordinary matter. While a huge majority of physicists believes that gravitational acceleration of particles and antiparticles is the same, there is room for surprises. The biggest surprise would be if gravitational acceleration of particles and antiparticles just differs in sign. Such a possible surprise (gravitational repulsion between matter and antimatter) is main assumption in our present paper.

In principle (as it was already known to Newton) we must distinguish between inertial mass $m_i$ and two gravitational masses (it may be better to say gravitational charges): the active gravitational mass which is the source of the gravitational field, and the passive gravitational mass which responds to an already existing field. As Newton correctly understood, because of the universality of the free fall (established by Galileo), inertial and passive gravitational mass can be considered as equal; while his third law (equality of action and reaction) implies that active and passive gravitational mass can be considered to be equal as well. So, the principle of equivalence of the inertial and gravitational mass of a body (known as the Weak Equivalence Principle) was born. This important principle has been used by Einstein as a cornerstone of general Relativity and so of modern Cosmology. Today, the WEP is the oldest and the most trusted principle of modern physics. However the WEP is tested only for ordinary matter. No one knows is it valid for antimatter, dark matter, dark energy, the eventually existing supersymmetric particles…

The existing experimental evidence and our assumption of gravitational repulsion between matter and antimatter may be summarized as:

$$m_i = m_g \; ; \; m_i = \bar{m}_i \; ; \; m_g + \bar{m}_g = 0 \qquad (6)$$

Here, as usually, a symbol with a bar denotes antiparticles; while indices *i* and *g* refer to inertial and gravitational mass. The first two relations in (1) are experimental evidence [5] while the third one is our assumption which dramatically differs from general conviction that $m_g - \bar{m}_g = 0$. It would be prudent to consider our assumption as valid for quarks and leptons and their antiparticles; and only on the basis of it to deduce gravitational mass (charge) of composite particles (for instance positronium, quarkonium and neutral pion must have zero gravitational mass; while negative pi meson, with quark structure $d\bar{u}$, must have the gravitational mass significantly smaller than the inertial one ).

Let's note that our conjecture (6) says nothing about gravitational proprieties of gauge and Higgs bosons. For instance, it is an open question if gravitational field can "see" difference between photons and antiphotons (currently considered as the same particle). Another uncertainty is the relation between active and passive gravitational mass, because in General Relativity, it is not fixed by anything like Newton's third law. For instance, we know that photons "feel" gravitational field, but even if we believe so, there is no experimental evidence that photons are also source of the gravitational field.

## 3. Major consequences of the conjecture

The usual statement that we live in the Universe completely dominated by matter is not true in the case of quantum vacuum, where (if the quantum field theories are right) virtual matter and antimatter "appear" in equal quantities. Thus, our hypothesis must have dramatic consequences for the quantum vacuum (a world of "virtual" matter) and only indirectly through it to our world of "real" matter. We limit ourselves to point out three major consequences of the conjecture (6).

First, it is immediately clear that a virtual particle-antiparticle pair is a system with zero gravitational mass and such a cancelation of gravitational masses might be important for an eventual solution of the cosmological constant problem. By the way, a similar cancellation of the opposite electric charges of particle and antiparticle in a virtual pair, leads to the zero density of the electric charge of the vacuum.

Second, a virtual pair may be considered as gravitational dipole with the gravitational dipole moment

$$\vec{p} = m\vec{d}; \quad p \approx m\lambdabar = \frac{\hbar}{c} \qquad (7)$$

Here, by definition, the vector $\vec{d}$ is directed from the antiparticle to the particle, and presents the distance between them. As the distance between particle and antiparticle is of the order of the Compton wavelength $\lambdabar = \hbar/mc$, we shall use the second of Equations (7) attributing to every virtual pair a gravitational dipole moment independent of mass.

The corresponding energy of the dipole in an external gravitational field characterized with acceleration $g_0$ is $\varepsilon = -\vec{p} \cdot \vec{g}_0$, i.e.

$$\varepsilon = -\frac{\hbar}{c} g_0; \quad \varepsilon = -\frac{\hbar}{c} \frac{GM_0}{r^2} \qquad (8)$$

The second of these equations concerns energy in the field of a spherical body with mass $M_0$.

Hence, polarization of the vacuum by a gravitational field might be possible. In order to grasp the difference between the polarization by an electric field and the eventual polarization by a gravitational field, let's remember that, as a consequence of polarization, the strength of an electric field is reduced in a dielectric. For instance, when a slab of dielectric is inserted into a parallel plate capacitor, the electric field between plates is reduced. The reduction is due to the fact that the electric charges of opposite sign attract each other. There is no reason to think about it in electrostatics, but let's note that if, instead of attraction, there was repulsion between charges with opposite sign, the electric field inside a dielectric would be augmented not reduced. But, according to our hypothesis, there is such repulsion between gravitational charges of different sign. Consequently, outside of a region in which a certain baryonic mass $M_0$ is confined, the eventual effect of polarization should be a gravitational field stronger than predicted by the Newton's law. The most important question is if the gravitational polarization of the vacuum can produce the same effect as presumed existence of dark matter. We will turn back to this question in section 4.

Third, the vacuum might permanently radiate. In order to understand it let's remember the illuminating example coming from Quantum Electrodynamics: creation of electron-positron pairs from the (Dirac) vacuum by an external (classical i.e. unquantized), constant and homogenous electrical field $E$.

In this particular case of the uniform electric field, the particle creation rate per unite volume and time is known [6] exactly:

$$\frac{dN_{e^+e^-}}{dt dV} = \frac{c}{\lambda_e^4} \left(\frac{E}{E_{cr}}\right)^2 \sum \frac{1}{n^2} \exp\left(-n\frac{E_{cr}}{E}\right); \quad E_{cr} = \pi \frac{m_e^2 c^3}{e\hbar} \qquad (9)$$

It is evident that particle creation rate is significant only for an electric field greater than the critical value $E_{cr}$.

The above phenomenon is due to both, the complex structure of the physical vacuum in QED and the existence of an external field. In the (Dirac) vacuum of QED, short-living "virtual" electron-positron pairs are continuously created and annihilated again by quantum fluctuations. A "virtual" pair can be converted into real electron-positron pair only in the presence of a strong external field, which can spatially separate electrons and positrons, by pushing them in opposite directions, as it does an electric field $E$. Thus, "virtual" pairs are spatially separated and converted into real pairs by the expenditure of the external field energy. For this to become possible, the potential energy has to vary by an amount $eE\Delta l > 2mc^2$ in the range of about one Compton wavelength $\Delta l = \hbar/mc$, which leads to the conclusion that the pair creation occurs only in a very strong external field $E$, greater than the critical value $E_{cr}$.

It is evident, that in the case of gravitational repulsion between matter and antimatter, a uniform gravitational field, just as a uniform electric field tends to separate "virtual" electrons and positrons, pushing them in opposite directions, which is a necessary condition for pair creation by an external field. But while an electric field can separate only charged particles, gravitation as a universal interaction may create particle-antiparticle pairs of both charged and neutral particles.

In the case of a uniform gravitational field, characterized with acceleration *g*, Equations (7) trivially transforms to

$$\frac{dN_{m\bar{m}}}{dtdV} = \frac{1}{\pi^2}\frac{g^2}{c^3}\frac{1}{\lambdabar_m^2}\sum_{n=1}^{\infty}\frac{1}{n^2}\exp\left(-n\pi\frac{c^2}{g\lambdabar_m}\right) \qquad (10)$$

with $\lambdabar_m$ being the reduced Compton wavelength.

Taking only the leading term $n=1$, distribution (10) has a maximum for

$$\lambdabar_{max} = \frac{\pi}{2}\frac{c^2}{g} \qquad (11)$$

Hence, in an external gravitational field *g* vacuum should radiate and the spectrum of radiation is dominated by $\lambdabar_{max}$. If so, the vacuum, as every radiating body, might be attributed a temperature *T*

$$kT = A\frac{\hbar c}{\lambdabar_{max}} = 4A\frac{1}{2\pi}\frac{\hbar}{c}g \qquad (12)$$

with *k* being the Boltzmann constant and *A*, a dimensionless constant which has to be determined.

The constant A in equations (12) can be determined using assumption that for all radial distances *r* vacuum radiates as a black body. If so, it must be, $\lambda_{max}T = b$, where *b* is Wien displacement law constant. From this condition it follows:

$$A = \frac{1}{2\pi}\frac{bk}{\hbar c} \approx \frac{1}{5} \qquad (13)$$

If the source of gravitation is a spherically symmetric body of the baryonic mass $M_b$, and if the acceleration $g_0$ is determined by the Newton law ($g_0 = GM_b/r^2$), the Eq.(12) gives:

$$kT = A\frac{\hbar c}{\pi}\frac{R_{S0}}{r^2}; \quad R_{S0} \equiv \frac{2GM_b}{c^2} \qquad (14)$$

Thus, we have attributed a temperature *T*, depending on the distance *r*, to the vacuum around a massive body. In the particular case $r = R_{S0}$, we rediscover the essential part of the famous Hawking's temperature of a black hole radiating as a black body. But this time Hawking's temperature seems to be just a particular case of a more general phenomenon. In fact Hawking's formula may be considered as a first hint, that in general, an external gravitational field vacuum acquires a "gravitational' temperature and consequently radiates. Namely, the mass of a black hole is collapsed, confined deep inside the horizon and surrounded by vacuum. There is nothing in the vicinity of the Schwarzschild radius, just vacuum. Hence, Hawking formula tells us that, under influence of a gravitational field, quantum vacuum in the vicinity of the Schwarzschild radius, radiates as a black body of temperature *T*. But, there is no special reason to limit vacuum radiation in the vicinity of the Schwarzschild radius; the phenomenon might exist for general *r*. By the way, formula (14) suggests that, deep inside the horizon of a black hole, temperature (and radiation) must be higher.

## 4. Toy model for vacuum polarization

As a toy model, let us consider quantum vacuum as a gas of independent gravitational dipoles in an external gravitational field, produced by a spherical body with baryonic mass $M_b$. Mathematically it is equivalent to the problem of independent electric dipoles in an external electric field *E*. The applied external field tends to orient dipoles in its direction, while the thermal agitation tends to randomize dipoles. Polarization (dipole moment per unite volume) is given by Langevin equation:

$$P = Np\left[\coth(x) - \frac{1}{x}\right] \qquad (15)$$

where $N$ is the number of dipoles per unite volume, p is the corresponding (gravitational or electric) dipole moment. For a gas of electric dipoles: $x \equiv pE/kT$. In the case of a gravitational dipole, following the Eq.(8), instead of $pE$, we may put $pg_0 = \hbar GM_b/c\,r^2$ ($g_0$ is the gravitational acceleration of the external gravitational field). It is not evident what to put instead of $kT$. The best what I can do is to start with the following fact: If the mass $M_b$ is surrounded by dark matter (or something producing the same effect) it follows from Eq.(2), that at large distances $r$, acceleration is dominated by a term

$$g_{dm} = \frac{G\rho_r}{r}, \quad \rho_r = \frac{dM_{dm}}{dr} \qquad (16)$$

where $\rho_r$ is a constant. Now, we may assume that $kT$ is proportional to the corresponding energy of the gravitational dipoles

$$kT \sim \frac{\hbar}{c} g_{dm} = \frac{\hbar}{c}\frac{G\rho_r}{r} \qquad (17)$$

It may be not a satisfactory solution but as we will see it works. Now, the polarization may be written as:

$$\vec{P} = -N\frac{\hbar}{c}\left[ctgh\left(\frac{M_b}{\rho_r r}\right) - \frac{\rho_r r}{M_b}\right]\vec{r}_0 \qquad (18)$$

where $\vec{r}_0$ is the corresponding unit vector.

It stays to guess what may be the value of N. We know that in our world of "real matter" dominant building blocks are quarks and leptons from the first generation. It seems reasonable to suppose that the same is true in the vacuum. If so, the dominant gravitational dipoles are virtual $\pi$ mesons. Thus N may be approximated by

$$N = \frac{1}{\lambda_\pi^3} = \left(\frac{m_\pi c}{2\pi\hbar}\right)^3 \qquad (19)$$

In full analogy with electrostatics, a gravitational mass density $\rho_p = -\nabla \cdot \vec{P}$ corresponds to the polarization (18). In the case of large $r$ (what is the main case of interest) elementary calculations lead to

$$\rho_p(r) = -\nabla \cdot \vec{P} = \frac{1}{3}N\frac{\hbar}{c}\frac{M_b}{\rho_r r^2} \qquad (20)$$

with corresponding mass inside sphere with radius $r$ equal to

$$M_p(r) = \frac{4\pi}{3}N\frac{\hbar}{c}\frac{M_b}{\rho_r}(r - R_0) \qquad (21)$$

where $R_0$ is a radius at which mass of dark matter can be considered to be zero.

A comparison of Eq. (21) with $M_{dm}(r) = \rho_r (r - R_0)$ immediately gives

$$\rho_r \equiv \frac{dM_{dm}}{dr} = B \frac{\sqrt{m_\pi M_b}}{\lambda_\pi} \qquad (22)$$

and

$$M_{dm}(r) = B \frac{\sqrt{m_\pi M_b}}{\lambda_\pi} (r - R_0) \qquad (23)$$

where $B$ is a dimensionless constant. Let's note that the above calculation leads to the numerical value $B = \sqrt{2/3} \approx 0.82$, but we may trust this number only as the accurate order of the magnitude.

The radial density profile (22) and the radial mass distribution (23) have the same form as given by Eq. (2) what are the essential characteristics of dark matter. This seems as a first indication that what we call dark matter eventually may be result of vacuum polarization and from now we make identification $M_{dm}(r) = M_p(r)$.

However, the most important fact is, that, in spite of the uncertainty about the value of $B$, Eq.(22) and Eq.(23) are very robust ones, in the sense that, the geometrical mean of the pion mass and total baryonic mass of a Galaxy, divided by Compton wavelength of a pion (i.e. $\sqrt{m_\pi M_b}/\lambda_\pi$) has the same order of magnitude as the observed radial dark matter density. For instance, in the case of a Galaxy like Milky Way, with a total baryonic mass estimated to be $M_b \approx 4 \times 10^{41} kg$ and viral radius of about 250kpc, equations (21) and (22) give respectively $\rho_r \approx 1.1 \times 10^{21} kg/m$ and $M_{dm} \approx 8.7 \times 10^{42} kg$ what is a few times larger than the observed values [8]. The other available observations lead to the same conclusion: in all cases $\sqrt{m_\pi M_b}/\lambda_\pi$ is an upper bound of the same order of magnitude as the exact value. In principle, observational data may serve to determine the appropriate value of $B$, but the trouble is that they are not very accurate; for instance, the halo's viral mass of our Galaxy [8] has not be constrained to better than a factor of $2-3$. It is amusing that (for a Galaxy) the choice

$$B = \frac{\Omega_b}{\Omega_{dm}} \approx 0.212 \quad or \quad B = \frac{\Omega_b}{\Omega_{dm}} = \frac{2}{3\pi} = 0.21220659 \qquad (24)$$

(where $\Omega_b$ and $\Omega_{dm}$ are baryon density of the Universe and dark matter density of the Universe) leads to numerical results in good agreement with observations. The relations (24) need a comment. First, let's point out a fact which (as far as I know) was not noticed before. The ratio of the volume of a Euclidean sphere ($V_s = 4\pi R^3/3$) and a closed Universe ($V_U = 2\pi^2 R^3$) equals $2/3\pi$. My conjecture (the second of equations (24)) is that it is the exact ratio of baryon density and dark matter density of the Universe. There is a striking "coincidence" between numerical value of my conjecture and the observed ratio [9] given by the first of relations (24). It may be just a coincidence, but if the ratio $\Omega_b/\Omega_{dm}$ is really determined by geometry, my guess is that we will never find dark particles; simply because they do not exist, as a new, unknown form of matter-energy.

In principle $B$ defined in (24) for a Galaxy shouldn't be the same for the Universe. In the following text

$$B_G = \frac{2}{3} \frac{1}{\pi} \quad and \quad B_U = \frac{1}{\pi} \qquad (25)$$

are respectively used for a galaxy and for the Universe (see also the following section). Different $B$ for a Galaxy and Universe should be result of different geometries. As already noticed the ratio $2/3\pi$ relates volumes and hence $\sqrt[3]{2/3\pi}$ which is not very different from $2/3$ should relate linear sizes: what is motivation for our choice (25). Of course, slightly different choice of $B_G$ is possible; for instance choices $B_G = 0.212$ and $B_G = 0.25$ are indistinguishable with current accuracy of observations.

The main weakness of the above toy model is that it neglects morphology of a Galaxy. All baryonic matter is supposed to be inside a sphere with radius $R_0$, much smaller than the radius $r_h$ of the halo of dark matter. Of course it is an oversimplification; even if $R_0$ is identified with the radius of the visible part of a Galaxy, there is still a lot of baryonic matter in much larger galactic halo. Hence Eq.(22) and Eq.(23) should be a better approximation for large $r$. In fact, because of geometry of a closed Universe it may be that Eq.(22) and Eq.(23) are better approximation for the whole Universe than a single Galaxy, what is considered in the following section.

## 5. Mass of the Universe

It is amusing to apply Equations (22), (23) and (25) to the Universe. The key advantage is that in this case we may identify the size of the halo of dark matter with the size of the Universe. Let's remember that travelling a proper distance $d_P = \pi R$ through a curved three-space (described by Friedman-Robertson-Walker metrics with cosmological scale factor $R$) brings us to the other end of the Universe. Hence the linear size of the halo of dark matter in the Universe may be identified with the maximum proper distance $d_P = \pi R$. The main difference between the case of a Galaxy and the Universe is that according to the cosmological principle every point of the Universe can be considered as its center, while a Galaxy has a single central point. A simple calculation (with $R_0 = 0$) leads to:

$$M_{dmU} = \frac{\Omega_b}{\Omega_{dm}} \left(\frac{R}{\lambda_\pi}\right)^2 m_\pi \approx 9.5 \times 10^{53} kg \tag{26}$$

$$M_{bU} = \left(\frac{\Omega_b}{\Omega_{dm}}\right)^2 \left(\frac{R}{\lambda_\pi}\right)^2 m_\pi \approx 2 \times 10^{53} kg \tag{27}$$

Here $M_{dmU}$ is mass of dark matter in the Universe; $M_{bU}$ is mass of baryons in the Universe and $R$ is cosmological scale factor entering the familiar Friedman equation

$$\frac{c^2}{H^2 R^2} = \Omega - 1 \tag{28}$$

Of course the above numbers can't be confronted with observations and we need estimates for the mass of the "visible" Universe. Let's note that the "radius" of the visible Universe is estimated to $R^{vis} = 14 Gpc$ and let's remember that mass of dark matter is proportional to the radius. Hence the mass of dark matter in the observable Universe should be a fraction ($\pi R/R^{vis}$) of the mass determined by Eq.(26), i.e.

$$M_{dmU}^{vis} = \frac{R^{vis}}{\pi R} \frac{\Omega_b}{\Omega_{dm}} \left(\frac{R}{\lambda_\pi}\right)^2 m_\pi \approx 1.1 \times 10^{53} kg \tag{29}$$

with the corresponding baryonic mass of the visible Universe

$$M_{bU}^{vis} = \frac{\Omega_b}{\Omega_{dm}} M_{dmU}^{vis} \approx 2.3 \times 10^{52} kg \qquad (30)$$

These numbers are very close to the existing speculations.

## 6. Mass of a Galaxy

Let's suppose that we know radius $r_h$ of a galactic halo. Then Equations (22), (23) and (25) lead to the following expressions for the baryonic mass ($M_{bG}$) and mass of dark matter ($M_{dmG}$) in a Galaxy.

$$M_{bG} = \left(\frac{\Omega_m}{\Omega_{dm}}\right)^4 \left(\frac{r_h}{\lambda_\pi}\right)^2 m_\pi \qquad (31)$$

$$M_{dmG} = \left(\frac{\Omega_m}{\Omega_{dm}}\right)^3 \left(\frac{r_h}{\lambda_\pi}\right)^2 m_\pi \qquad (32)$$

Unfortunately, while the "radius" of the Universe, i.e. the scale factor determined by Eq.(28), is relatively well known, the radius of galactic halo is poorly known even in the case of our Galaxy; for Milky Way a crude approximation is $r_h \approx 250 kpc$. If so, the Eq.(31) and Eq.(32) give for the Milky Way:

$$M_{bG} \approx 3.8 \times 10^{41} kg ; \; M_{dmG} \approx 1.8 \times 10^{42} kg ; \; M_{tot} \approx 2.2 \times 10^{42} kg \qquad (33)$$

what is in very good agreement with observations (See [7] and References therein)

## 7. Size of a Galactic Halo

As already noticed, the size of a galactic halo is not a well known quantity. In order to find a crude approximation, let's denote by $N_G$ the total number of galaxies in the Universe and for simplicity let's assume that they all have the same mass $M_{bG}$ and $M_{dmG}$ so that we may write $M_{dmU} = N_G M_{dmG}$. This simple relation together with Equations (26) and (32) lead to a nice approximation:

$$r_h = \frac{\Omega_{dm}}{\Omega_b} \frac{R}{\sqrt{N_G}} \qquad (34)$$

For instance $N_G = 4 \times 10^{11}$ and $N_G = 6 \times 10^{11}$ give respectively $r_h \approx 290 kpc$ and $r_h \approx 235 kpc$.

By the way, if a typical Galaxy has baryonic mass $3.8 \times 10^{41} kg$ (as estimated in (33)), than from Eq.(27) follows $N_G \approx 5.3 \times 10^{11}$ corresponding to $r_h \approx 252 kpc$!

## 8. Comments

The near future will show if the present paper is useless or significant. But for sure this paper may eventually be important only because of the ideas presented herein, while the extremely simplified considerations must be promptly replaced by more rigorous calculations. However, in spite of simplifications, there is an impressive numerical agreement of our results [(29), (30), (33) and section 7) with the basic observational data.

The key relation (22) may be considered as a starting point and has its own value independently of speculations leading to it. Even if there is no gravitational repulsion between matter and antimatter, relation (24) is an interesting fact which must be understood.

In conclusion, I think we must seriously take the idea that dark matter and dark energy are not new, unknown forms of matter-energy but an effect of complex interaction between the quantum vacuum and known baryonic matter.

**Dedication**

This paper is dedicated to my father Slavko, and my children Ivan and Anja-Milica